\documentclass[aps,prb,twocolumn,amsmath,amssymb,superscriptaddress,floatfix]{revtex4}
\usepackage{graphicx} 
\usepackage{bm}
\usepackage[usenames]{color}
\bibstyle{apsrev.bib}

\newcommand{\be}{\begin{equation}}
\newcommand{\ee}{\end{equation}}
\newcommand{\beqn}{\begin{eqnarray}}
\newcommand{\eeqn}{\end{eqnarray}}

\begin{document}

\title{Exact bounds on the energy gap of transverse-field Ising chains by mapping to random walks}

\author{R\'obert Juh\'asz}
\email{juhasz.robert@wigner.hu}
\affiliation{Wigner Research Centre for Physics, Institute for Solid State Physics and Optics, H-1525 Budapest, P.O.Box 49, Hungary}

\date{\today}

\begin{abstract}
Based on a relationship with continuous-time random walks discovered by Igl\'oi, Turban, and Rieger [Phys. Rev. E {\bf 59}, 1465 (1999)], we derive exact lower and upper bounds on the lowest energy gap of open transverse-field Ising chains, which are explicit in the parameters and are generally valid for arbitrary sets of possibly random couplings and fields.
In the homogeneous chain and in the random chain with uncorrelated parameters, both the lower and upper bounds are found to show the same finite-size scaling in the ferromagnetic phase and at the critical point, demonstrating the ability of these bounds to infer the correct finite-size scaling of the critical gap. 
Applying the bounds to random transverse-field Ising chains with coupling-field correlations, a model which is relevant for adiabatic quantum computing, the finite-size scaling of the gap is shown to be related to that of sums of independent random variables. We determine the critical dynamical 
exponent of the model and reveal the existence of logarithmic corrections at special points.   
\end{abstract}

\maketitle

\section{Introduction}

There is a class of one-dimensional quantum lattice models which have in common that the excitation energies are given by the eigenvalues of certain tridiagonal matrices. The most prominent example is the transverse-field Ising chain (TFIC), which can be mapped to a fermion chain with quadratic terms by the well-known Jordan-Wigner transformation \cite{pfeuty}. A closely related model is the spin-$\frac{1}{2}$ XY chain \cite{lsm}, which can be mapped to two independent TFIC-s \cite{perk,fisherXX,ijr}. Another representatives of this class are the fermionic hopping models on a one-dimensional lattice.     
In this paper, we focus on the energy gap between the ground state and the first excited state of TFIC. The relevance of studying the gap is given by the existence of a continuous phase transition of the model from a paramagnetic to a ferromagnetic phase when the the strength of the transverse field is decreased. In the ferromagnetic phase, the energy gap closes exponentially with the system size $L$ and the ground state and first excited state become asymptotically degenerate, showing the spontaneous symmetry breaking of the infinite system. In the critical point, the gap closing is slower than exponential, in general a power law, $\epsilon\sim L^{-z}$, where $z$ is the critical dynamical exponent, except of the TFIC with uncorrelated random \cite{fisher} or certain aperiodically modulated \cite{luck,ikr} couplings, in which the gap vanishes according to a stretched exponential law $\epsilon\sim e^{-const\cdot L^{\Psi}}$ with $\Psi<1$.   
The paramagnetic phase is gapped in general, but the uncorrelated random model is an exception also in this respect: in the Griffiths-McCoy phase, the gap vanishes algebraically with a non-universal dynamical exponent depending on the control parameter \cite{griffiths,mccoy,ir_pre_98,ijl}.    
Recently, the interest in the scaling of energy gap has increased from the side of adiabatic quantum computing \cite{albash,hauke}. Due to its solvability in polynomial time, the TFIC is an ideal testing ground for different quantum annealing protocols in which the TFIC is slowly driven from a large initial transverse field through the critical point to a classical target Hamiltonian with a zero transverse field \cite{kadowaki,dziarmaga,caneva,dziarmaga_rams,lucas,rams,delcampo}. A crucial difficulty of quantum annealing is the breaking of adiabaticity at finite annealing rates, i.e. the system will be excited from the instantaneous ground state to higher-lying states during the procedure so that the end state may contain defects with some probability \cite{kibble,zurek,lukin,king}. The rate of formation of defects which are to be avoided from the point of view of adiabatic quantum computing is more enhanced at small instantaneous gaps: according to the adiabatic theorem \cite{amin,albash}, the necessary computation time is the maximum of the transition matrix element of the time derivative of the instantaneous Hamiltonian divided by the square of the instantaneous gap. The knowledge of the gap of the static TFIC, especially the minimal gap experienced during the annealing procedure is thus important for estimating the efficiency of the protocol or for devising an optimal protocol which minimizes the probability of defect formation. We note here that, due to the parity symmetry of the TFIC, the gap which is relevant for quantum annealing is the gap between the ground state and the first excited state within the ground-state sector, the other sector being unavailable for the dynamics. Nevertheless, the finite-size scaling of this energy difference at the critical point is in general similar to that of the lowest gap.     
     
Apart from the homogeneous chain, the eigenvalue problem of which is analytically solvable \cite{lsm,pfeuty,dr}, there does not exist a closed form of the lowest gap for a general set of couplings and transverse fields. 
Therefore various numerical methods and analytic approximations have been developed to estimate the gap.
For the TFIC with uncorrelated random parameters, the known results on the energy gap are obtained mainly by the strong-disorder renormalization group (SDRG) method \cite{mdh,fisher,young,jli,im} and by numerical diagonalization \cite{yr}. 
In addition to this, there exists an approximative formula for the gap of the open chain obtained in Refs. \cite{itksz,ir98}, which is accurate in the ferromagnetic phase and yields the correct finite-size scaling of the gap even at the critical point. This formula is similar to that obtained perturbatively for the periodic chain in Ref. \cite{luck}. 
In Ref. \cite{knysh}, the terms of the characteristic polynomial beyond the second order one were neglected and the resulting quadratic equation was solved to estimate the scaling of the gap of the random TFIC to the second excited state. 
Ref. \cite{alcaraz} went further in this direction and used the truncated characteristic polynomial of much higher order to obtain very accurate numerical estimates of the lowest gap at the critical point and in the ferromagnetic phase. 
Furthermore, in Ref. \cite{alcaraz} the Laguerre bound for the smallest root of the characteristic polynomial, which can be calculated from the first three coefficients of the characteristic polynomial, was studied and found to correctly reproduce the finite-size scaling of the exact gap at the critical point and in gapless phases. 

In this paper, we extend the results concerning the gap of the open TFIC by providing exact lower and upper bounds which are generally valid for any set of possibly random couplings and fields, and are explicit in the parameters of the model. Due to this latter property, they are promising starting points of possible analytic treatments toward the determination of the finite-size scaling of the gap. We will demonstrate the power of these bounds by obtaining the finite-size scaling of the gap of the random TFIC with coupling-field correlations, which is relevant in the context of adiabatic quantum computing and has been studied by several authors \cite{binosi,hoyos_epl,getelina,gh,knysh,shirai}. 
The derivation rests on an exact relationship between the open TFIC and continuous-time random walks with an absorbing boundary. Some elements of this relationship were discovered in Ref. \cite{ir_pre_98} but, in the most complete form, it was formulated by Igl\'oi, Turban, and Rieger in Ref. \cite{itr}.    
In addition to this, the derivation uses properties of the quasistationary distribution of Markov processes and known expressions of the mean time to absorption.

The paper is organized as follows. In section \ref{route}, we recapitulate the bases of the derivation of the bounds: the calculation of excitation spectrum of the TFIC, the mapping to a Markov process, and the relation of the spectral gap to the mean time to absorption. The lower and upper bounds are derived in sections \ref{lb} and \ref{ub}. Next, these bounds are applied to the homogeneous chain and the random chain with correlated and uncorrelated randomness. Finally, results are discussed in section \ref{discussion}. Some of calculations are presented in the Appendix.

\section{The route from TFIC to Markov chains}
\label{route}

\subsection{Excitation spectrum}
\label{ex}

We consider the transverse-field Ising chain with $L>1$ spins and with open boundary condition:
\be
{\cal H}=
-\sum_{n=1}^{L-1}\frac{J_n}{2}\sigma_n^x \sigma_{n+1}^x-\sum_{n=1}^L\frac{h_n}{2}\sigma_n^z,
\label{H}
\ee
where $\sigma_n^x$ and $\sigma_n^z$ are Pauli operators at site $n$, and the couplings $J_n$ and external fields $h_n$ are assumed to be non-zero unless stated otherwise. 
 
By the Jordan-Wigner transformation, 
\beqn
c_n^{\dagger}+c_n=(\prod_{m<n}-\sigma_m^z)\sigma_n^x, \nonumber \\ 
c_n^{\dagger}-c_n=i(\prod_{m<n}-\sigma_m^z)\sigma_n^y, \nonumber \\
n=1,2,\dots,L
\eeqn
the Hamiltonian in Eq. (\ref{H}) can be written in a quadratic form of fermion creation
($c_n^{\dagger}$) and annihilation ($c_n$) operators \cite{lsm,pfeuty}:
\be
{\cal H}=
-\sum_{n=1}^{L-1}\frac{J_n}{2}(c_n^{\dagger}-c_n)(c_{n+1}^{\dagger}+c_{n+1})
 -\sum_{n=1}^Lh_n(c_n^{\dagger}c_n-\frac{1}{2}).
\label{Hfermion}
\ee
A subsequent Bogoliubov-Valatin transformation 
\be
\eta_k=\sum_{n=1}^L\left[\frac{\phi_{kn}+\psi_{kn}}{2}c_n + \frac{\phi_{kn}-\psi_{kn}}{2}c_n^{\dagger}\right]  \quad k=1,\dots,L
\ee
with appropriately chosen coefficients $\phi_{kn}$ and $\psi_{kn}$ brings the Hamiltonian in Eq. (\ref{Hfermion}) to a diagonal form
\be
{\cal H}=\sum_{k=1}^{L}\epsilon_k(\eta_k^{\dagger}\eta_k-\frac{1}{2}).
\ee
The ground state of the model is the vacuum state of $\eta_k$ fermions, and the positive excitation energies $\epsilon_k$ are obtained as the $\pm\epsilon_k$ eigenvalue pairs of the symmetric, tridiagonal matrix 
\be 
H=
\begin{pmatrix}
0   & h_1&               &        &               &    & &\cr
h_1   &0            & J_1 &        &               &    & &\cr
    & J_1          &  0           & h_2 &         &    & &\cr  
    &               &  h_2         & 0     & \ddots        &    & &\cr
&   &               &  \ddots        & \ddots     &         &    &\cr
& &    &               &              & 0 & J_{L-1}       &  \cr
& &    &               &               &   J_{L-1}   &    0    & h_L \cr  
& &    &               &               &        &  h_L          & 0 
\end{pmatrix}.
\label{Tmatrix}
\ee
Equivalently, the excitation energies $\epsilon_k$ are the singular values of the bidiagonal matrix \cite{juhasz2022}
\be 
M=
\begin{pmatrix}
h_1   & &               &        &               &     \cr
J_1   & h_2            &  &        &               &     \cr
    & J_2          &  \ddots           &  &         &     \cr  
    &               & \ddots         &      &         &     \cr
    &               &               &   J_{L-2}   &    h_{L-1}    &  \cr  
    &               &               &        &  J_{L-1}          & h_L 
\end{pmatrix},
\label{bidiag}
\ee
or the squared excitation energies $\epsilon_k^2$ are the eigenvalues of the symmetric, tridiagonal matrix
\beqn 
&MM^T=& \nonumber \\
&=\begin{pmatrix}
h_1^2   & h_1J_1&               &        &                   \cr
h_1J_1   & h_2^2+J_1^2            & h_2J_2 &        &                  \cr
    & h_2J_2          &     h_3^2+J_2^2        &  \ddots &            \cr  
    &               &    \ddots       &  \ddots   &         &    \cr
              &               &      &        & h_{L-1}J_{L-1} \cr  
            &               &        &  h_{L-1}J_{L-1}          & h_L^2+J_{L-1}^2 
\end{pmatrix}.& \nonumber \\
\label{MMT}
\eeqn
We note that, instead of $MM^T$ one can use $M^TM$ equally well, since they are similar matrices, and the latter is obtained from the former (in a reversed order of rows and columns) by the replacements 
\be
h_i\leftrightarrow h_{L-i+1}, \quad J_i\leftrightarrow J_{L-i}, 
\label{inversion}
\ee
which amounts to an inversion of the original model.

\subsection{Mapping to a Markov chain}
\label{mapping}

The essence of the relationship revealed in Ref. \cite{itr} is that the positive matrix $MM^T$ given in Eq. (\ref{MMT}) is similar to the transient part of the rate matrix of a stochastic process (up to a global minus sign). 
To be more concrete, let us introduce the diagonal matrix $S={\rm diag}\{\alpha_1,\alpha_2,\dots,\alpha_L\}$ with elements $\alpha_1=1$, $\alpha_n=\prod_{i=1}^{n-1}\left(-\frac{h_i}{J_i}\right)$, $n=2,\dots,L$, by which the matrix $MM^T$ can be transformed to 
\beqn 
&T\equiv -S^{-1}MM^TS =&  \nonumber \\
&=\begin{pmatrix}
-h_1^2   & h_1^2&               &        &                   \cr
J_1^2   & -h_2^2-J_1^2            & h_2^2 &        &                  \cr
    & J_2^2          &     -h_3^2-J_2^2        &  \ddots &            \cr  
    &               &    \ddots       &  \ddots   &         &    \cr
              &               &      &        & h_{L-1}^2 \cr  
            &               &        &  J_{L-1}^2          & -h_L^2-J_{L-1}^2 
\end{pmatrix}.& \nonumber \\
\label{-T}
\eeqn
One can see that the non-diagonal elements of this matrix are nonnegative, and the sums of the elements in all but the last row are zero. 
Extending $T$ with the $L$-component column vector ${\bf a}=(0,0,\dots,h_L^2)^T$ and $L$-component row vector ${\bf 0}=(0,0,\dots,0)$ in the form 
\be 
Q=
\begin{pmatrix}
T  & {\bf a}     \cr
{\bf 0}    &  0             
\end{pmatrix},
\label{Q}
\ee 
one obtains a stochastic matrix $Q$ of order $L+1$. 
This can be interpreted as the infinitesimal generator of a continuous-time Markov process (random walk) on the states labeled by $1,2,\dots,L+1$, with transition rates 
$Q_{n,n+1}=h_n^2$ for $n=1,\dots,L$ and $Q_{n+1,n}=J_n^2$ for $n=1,\dots,L-1$. Thus state $L+1$, having a zero exit rate, is an absorbing state. 
The matrix $Q$ has a zero eigenvalue $\lambda_0=0$ which corresponds to the (absorbing) stationary state ($p_{L+1}=1$, $p_n=0$ for $n=1,\dots,L$), while all other eigenvalues are negative and given by $-\lambda_k=-\epsilon_k^2$ due to the similarity of $-T$ and $MM^T$.    

\subsection{Mean time to absorption and the spectral gap}
\label{mfpt}

The key point of formulating bounds on the top eigenvalue $-\lambda_1=-\epsilon_1^2$ is its relation with the mean time to absorption (also known as first-passage time\cite{redner}) when the equivalent stochastic process is initiated in its quasistationary distribution.
To see this, we use well-known properties of irreducible continuous-time Markov chains with a finite number of states \cite{darroch,doorn}. 
In this case, the top eigenvalue $-\lambda_1$ is unique, simple and, due to the similarity to a symmetric matrix, it is real. 
Moreover, the associated left eigenvector of $T$ can be chosen to be componentwise positive. 
A distribution $q=(q_1,q_2,\dots,q_L)$ is called quasistationary if it remains constant under the condition of non-absorption when $q$ is the initial distribution. Finite, irreducible Markov chains are known to have a unique quasistationary distribution which is the (unique) left eigenvector of $T$ associated with eigenvalue $\lambda_1$: $qT=-\lambda_1q$, normalized as $\sum_{n=1}^Lq_i=1$. 
One can then show that the row vector with $L+1$ components 
$q=(q_1,q_2,\dots,q_L,-1)$ is a left eigenvector of $Q$: $qQ=-\lambda_1q$. 
Considering the distribution $P(t=0)=q+s=(q_1,q_2,\dots,q_L,0)$, where $s=(0,0,\dots,0,1)$ denotes the stationary distribution (left eigenvector of $Q$ with zero eigenvalue) as an initial distribution of the master equation $\frac{dP(t)}{dt}=P(t)Q$, we obtain for the time evolution
\be
P(t)=P(0)e^{Qt}=s+e^{-\lambda_1t}q.
\ee
The probability of not being absorbed on site $L+1$ up to time $t$ is then 
$1-P_{L+1}(t)=e^{-\lambda_1t}$ and the mean time to absorption (mean first-passage time) is 
\be 
\tau_{\rm qs}=\frac{1}{\lambda_1}.
\label{taulambda}
\ee
Here the subscript $qs$ refers to that the initial distribution was the quasistationary one.            

\section{Lower bound on the gap}
\label{lb}

The mean time to absorption when the process starts from site $n$, denoted by $\tau_n$ can be analytically calculated for the Markov process defined by $Q$. As it is well-known, the mean times to absorption $\tau_1,\tau_2,\dots,\tau_L$ satisfy the so called backward master equations \cite{vankampen}, which can be written in the compact form
\be 
T\tau=-{\bf 1},
\label{back}
\ee
with the column vectors $\tau=(\tau_1,\tau_2,\dots,\tau_L)^T$ and ${\bf 1}=(1,1,\dots,1)^T$. These can be solved to give \footnote{For an interesting analogy between the solution of Eqs. (\ref{back}) and the SDRG procedure, see Ref. \cite{mg}}:   
\be
\tau_n=\sum_{l=n}^L\sum_{m=1}^l\frac{1}{h_l^2}\prod_{i=m}^{l-1}\frac{J_i^2}{h_i^2}
\label{tau}
\ee
with the convention that the product is $1$ whenever $m=l$.
Obviously, the mean times to absorption monotonically increase with decreasing indices $n$ (being farther from the absorbing site):
\be
\tau_m<\tau_n \quad {\rm if} \quad  m>n.
\label{mon}
\ee 
Since $\tau_{qs}=\sum_{n=1}^Lq_n\tau_m$, $q_n$ denoting the quasistationary distribution as before, we obtain immediately that 
\be
\tau_{qs}<\tau_1.
\ee
Thus, by Eq. (\ref{taulambda}) we have the following lower bound on the gap
\be
\epsilon_1>\frac{1}{\sqrt{\tau_1}}=\left[\sum_{l=1}^L\sum_{m=1}^l\frac{1}{h_l^2}\prod_{i=m}^{l-1}\frac{J_i^2}{h_i^2}\right]^{-\frac{1}{2}}.
\ee
This bound turns out to be closely related to the coefficients of the characteristic polynomial of $MM^T$, $P(\lambda)=\det(MM^T-\lambda\mathbb{I})=\sum_{n=0}^LC_n\lambda^n$. The constant term is $C_0=\det(MM^T)=[\det(M)]^2=h_1^2h_2^2\cdots h_L^2$, where we used Eq. (\ref{bidiag}).  
The coefficient of the linear term, as it is described in Appendix \ref{C1}, can be shown to be 
\be
C_1=-\tau_1C_0.
\label{c1tau1}
\ee
On the other hand, from the factorized form of the characteristic polynomial 
$P(\lambda)=\prod_{n=1}^L(\lambda_n-\lambda)$ it is clear that $C_1=-C_0\sum_{n=1}^L\lambda_n^{-1}$ (which is one of Vi\`ete's formulae), therefore $\tau_1$ is simply the sum of reciprocal eigenvalues:
\be
\tau_1=\sum_{n=1}^L\lambda_n^{-1}=\sum_{n=1}^L\epsilon_n^{-2}.
\ee

Now we compare this bound to Laguerre's lower bound $\lambda_{\rm Lb}$ of all roots used in Ref. \cite{alcaraz}, which is composed of the first three coefficients of the characteristic polynomial as
\be 
\frac{1}{\lambda_{Lb}}=-\frac{1}{L}\frac{C_1}{C_0}+ \frac{\sqrt{L-1}}{L}
\sqrt{(L-1)\left(\frac{C_1}{C_0}\right)^2-2L\frac{C_2}{C_0}}.
\label{Lb}
\ee
One can notice, that formally substituting $C_2=0$ in the Eq. (\ref{Lb}) leads to $1/\lambda_{Lb}(C_2=0)=\tau_1$. Since, according to Vi\`ete's formulae 
$\frac{C_2}{C_0}=\sum_{i<j}\frac{1}{\lambda_i\lambda_j}$, which is positive, the lower bound obtained by the random-walk mapping is less sharp than Laguerre's lower bound:
\be
\frac{1}{\tau_1}<\lambda_{Lb}<\lambda_1.
\ee
Nevertheless, we will see later that even this weaker bound which, on the other hand, has the advantage of being simpler than Laguerre's bound, is sufficient to infer the finite-size scaling of the gap at the critical point. 

We close this section with the comparison of the approximate formula for the gap derived in Refs. \cite{itksz,ir98} to our lower bound. As it is shown in Appendix \ref{app:comparison}, the approximate gap is below the lower bound, $\epsilon_{\rm app}<\frac{1}{\sqrt{\tau_1}}<\epsilon_1$, so the latter is a better approximation of the exact gap. 

\section{Upper bound}
\label{ub}

The relationship with random walks described in the previous sections also enables us to establish upper bounds on the gap. 
We can obtain an upper bound immediately from the monotonicity of $\tau_n$, which implies $\tau_{\rm qs}>\tau_L$, yielding the upper bound 
\be
\epsilon_1<\frac{1}{\sqrt{\tau_L}}=
\sum_{m=1}^L\frac{1}{h_L^2}\prod_{i=m}^{L-1}\frac{J_i^2}{h_i^2}.
\ee
One can improve this bound by noting that using $M^TM$ rather than $MM^T$ leads to a different bound $\overline{\tau}_L$, which is related to $\tau_{L}$ through the inversion in Eq. (\ref{inversion}). Then $\epsilon_1<1/\tau_L^{\rm max}$, where 
$\tau_L^{\rm max}=\max\{\tau_L,\overline{\tau_L}\}$. 
Yet, this upper bound is not sufficiently sharp to have the same finite-size scaling as the exact gap at the critical point in general. 

Nevertheless, this requirement can be fulfilled with a sharper bound constructed by the help of the stationary distribution of a modified Markov chain. 
Let us consider the Markov process as before but with the restriction to sites $1,2,\dots,L$ and with $h_L=0$. The corresponding rate matrix, which is obtained from $T$ by setting $h_L=0$, will be denoted by $T'$. This process has a non-trivial stationary state $(p_1,p_2,\dots,p_L)$, which is the left eigenvector of $T'$ associated with the zero eigenvalue. By recursion, we obtain this distribution in the form
\be
p_n=\frac{\prod_{i=1}^{n-1}\frac{h_i^2}{J_i^2}}{\sum_{n=1}^L\prod_{i=1}^{n-1}\frac{h_i^2}{J_i^2}}, \quad n=1,2,\dots,L.
\label{stac}
\ee
Comparing the quasistationary distribution to this one, one has the intuition that the steady loss of probability at site $L$ in the former case leads to a depletion of probabilities near the absorbing site in favor of those near the opposite end of the chain (keep in mind that $q$ is a normalized distribution). 
Indeed, as it is proved in Appendix \ref{major}, there exists an index $1<n^*<L$ such that 
\beqn
q_n&>&p_n \quad {\rm if} \quad n<n^*, \nonumber \\
q_{n^*}&\ge&p_{n^*}, \quad {\rm and} \nonumber \\
q_n&<&p_n \quad {\rm if} \quad n>n^*. 
\label{pq}
\eeqn
This implies, together with the monotonicity of $\tau_n$ (see Appendix \ref{major}) that the mean time to absorption in the original process starting from the stationary distribution of the modified process
\be
\tau_s=\sum_{n=1}^Lp_n\tau_n
\label{taus}
\ee  
fulfills the inequality 
\be
\tau_s<\tau_{qs}. 
\label{sqs}
\ee
We obtain then the following upper bound on the gap:
\beqn
&\epsilon_1<\displaystyle\frac{1}{\sqrt{\tau_s}}=& \nonumber \\
&=\left[\displaystyle\sum_{n=1}^L\prod_{i=1}^{n-1}\frac{h_i^2}{J_i^2}\right]^{\frac{1}{2}}
\left[
\displaystyle\sum_{n=1}^L\left(\prod_{i=1}^{n-1}\frac{h_i^2}{J_i^2}\right)\displaystyle\sum_{l=n}^L\sum_{m=1}^l\frac{1}{h_l^2}\prod_{j=m}^{l-1}\frac{J_j^2}{h_j^2}\right]^{-\frac{1}{2}}.& \nonumber \\
\label{upper}
\eeqn
Note that, as opposed to the lower bound $1/\sqrt{\tau_1}$, the upper bound in Eq. (\ref{upper}) does not show the inversion symmetry, therefore performing the replacement given in Eq. (\ref{inversion}) yields a different upper bound, $1/\sqrt{\overline{\tau}_s}$. We have then 
$\epsilon_1<1/\sqrt{\tau^{\rm max}_s}$, where $\tau_s^{\rm max}=\max\{\tau_s,\overline{\tau}_s\}$.   

\section{Application of the bounds}
\label{app}

\subsection{Homogeneous chain}

First, we will test the bounds obtained in the previous sections for the homogeneous transverse-field Ising chain with couplings $J_n=J>0$ and fields $h_n=h>0$. Then the summations in Eq. (\ref{tau}) can be performed and, introducing the ratio $r=J^2/h^2$, we obtain
\be 
\tau_n=\frac{1}{h^2}\frac{(L-n+1)(1-r)+r^{L+1}-r^n}{(1-r)^2}
\label{tau_h}
\ee
if $r\neq 1$, while for $r=1$, i.e. at the critical point we find
\be
\tau_n=\frac{1}{2h^2}(L-n+1)(L+n).
\ee  
Performing the summation in Eq. (\ref{taus}) with 
$p_n=\frac{1-r}{1-r^L}r^{L-n}$ results ultimately in 
\be 
\tau_s=\frac{1}{h^2}\frac{1-r^{2L+1}+(2L+1)r^L(r-1)}{(1-r^L)(1-r)^2}
\label{tau_s_h}
\ee
for $r\neq 1$, while, for $r=1$ we find 
\be
\tau_s=\frac{1}{h^2}\left(\frac{L^2}{3}+\frac{L}{2}+\frac{1}{6}\right).
\ee

Let us now check the validity of the lower and upper bounds in the different phases of the model.
In the ferromagnetic phase ($r<1$), the gap closes exponentially with the system size in leading order as\cite{pfeuty} $\epsilon_1\simeq J\left(\frac{h}{J}\right)^L$.
The lower bound in this phase is $\frac{1}{\sqrt{\tau_1}}\simeq (J-\frac{h^2}{J})(\frac{h}{J})^L$, thus the prefactor is smaller than the exact one, while the leading term of the upper bound agrees with that of the exact gap: $\frac{1}{\sqrt{\tau_s}}\simeq J(\frac{h}{J})^L$.    

In the paramagnetic phase ($r>1$), the gap is non-zero in the limit $L\to\infty$: $\epsilon_1=h-J$. The lower bound is vanishing with $L$ as $\frac{1}{\sqrt{\tau_1}}\simeq \frac{\sqrt{h^2-J^2}}{\sqrt{L}}$, thus it is not useful here, whereas the upper bound is asymptotically constant: $\lim_{L\to\infty}\frac{1}{\sqrt{\tau_s}}=(h-J)(1+\frac{J}{h})$.    

At the critical point, $J=h=1$, the gap vanishes with the system size as\cite{pfeuty} $\epsilon_1=2\sin(\frac{\pi}{2}\frac{1}{2L+1})=\frac{\pi}{2}\frac{1}{L}+O(L^{-2})$. For the lower and upper bounds we find here in leading order:  
\be 
\frac{1}{\sqrt{\tau_1}}\simeq \frac{\sqrt{2}}{L}, \qquad 
\frac{1}{\sqrt{\tau_s}}\simeq \frac{\sqrt{3}}{L}.
\ee
Thus, at the critical point both bounds show the same finite-size scaling, and the scaled gap in the large-$L$ limit $\lim_{L\to\infty}\epsilon_1L=\frac{\pi}{2}=1.57\dots$ is bounded relatively tightly by $\sqrt{2}=1.41\dots$ and $\sqrt{3}=1.73\cdots$.  
Finally, it is interesting to note that, for $J=h=1$,  $\tau_1$ is the sum of natural numbers up to $L$ , whereas $\tau_s$ is the sum of the squares of natural numbers up to $L$, divided by $L$. 

\subsection{Random chain with local coupling-field correlations}

Next, we apply the bounds obtained in the previous sections to infer the finite-size scaling of the gap of the random TFIC in which the fields are correlated with neighboring couplings. We consider a general form of such correlations used in Ref. \cite{shirai}: the couplings are independent, identically distributed random variables, while the fields are fixed by neighboring couplings as 
\beqn
h_1&=&J_1^{1-s},  \nonumber \\
h_n&=&J_{n-1}^sJ_n^{1-s}, \quad 1<n<L \nonumber \\
h_L&=&J_{L-1}^s.
\label{hJ}
\eeqn    
Here, the parameter $s$ is in the range $0\le s\le 1$. The special case $s=0$ was studied in Refs. \cite{binosi,hoyos_epl,getelina,gh}, while the symmetric case $s=\frac{1}{2}$ was considered in Ref. \cite{knysh}.
By the choice of the fields as given in Eq. (\ref{hJ}), the model is critical, and the fluctuations of the sample-dependent control parameter $u_L=\sum_{n=1}^{L-1}\ln(J_n/h_n)$, which follow the central limit theorem for uncorrelated randomness, become independent of $L$.  
The relevance of this for adiabatic quantum computing is that, for such a choice of fields (multiplied by a global driving field), the minimal gap during the annealing process, which closes stretched exponentially for uniform fields, will be less tiny, closing only algebraically with $L$.      
According to numerical results\cite{hoyos_epl,getelina} obtained with a power-law distribution of couplings 
\be
\rho(J)=\frac{1}{D}J^{-1+\frac{1}{D}}, 
\label{pl}
\ee
with the support $0<J<1$ and the parameter $D>0$ which controls the strength of disorder, the critical dynamical exponent is $z=1$ for weak enough disorder $D<D_c$, and $z>1$ otherwise. Later it was confirmed by an exact lower bound on the dynamical exponent \cite{shirai}. 

We will now show that the lower and the upper bound derived in this paper show the same typical finite-size dependence and determine thereby the dynamical exponent of the model. 
Expressing the fields with the couplings as given in Eq. (\ref{hJ}), we obtain for the sums relevant for the lower and upper bounds:
\be
\tau_1=\sum_{l=1}^L\sum_{m=1}^lJ_l^{-2(1-s)}J_{m-1}^{-2s}
\label{tau1c}
\ee 
and 
\be
\tau_s=\left[\sum_{n=1}^LJ_n^{-2s}\right]^{-1}
\sum_{n=1}^L\sum_{l=n}^L\sum_{m=1}^lJ_{n-1}^{-2s}J_{l}^{-2(1-s)}J_{m-1}^{-2s},
\label{tausc}
\ee
with the convention $J_0=J_L=1$. 
To shorten the notation we introduce $x_n\equiv J_n^{-2}$. Extending the upper limit of the second sum in Eq. (\ref{tau1c}) to $L$, we have 
\be
\tau_1<\tau_1'=\left[\sum_{m=1}^Lx_m^{s}\right]\left[\sum_{l=1}^Lx_l^{1-s}\right],
\label{tau1prime}
\ee 
and we obtain thereby another lower bound $1/\sqrt{\tau_1'}<\epsilon_1$, which is less sharp than the original one but more appropriate for our purposes. 

Considering the upper bound, we rewrite Eq. (\ref{tausc}) as 
$\tau_s=\left[\sum_{n=1}^Lx_n^{s}\right]^{-1}\sum_{l=1}^L\sum_{n=1}^l\sum_{m=1}^lx_l^{1-s}x_{n-1}^sx_{m-1}^s$. By restricting the lower limit of the first sum and the upper limits of the remaining two sums of the triple sum to $[L/2]+1$, where $[L/2]$ denotes the integer part of $L/2$, we have 
\be
\tau_s>\tau_s'=\left[\sum_{n=1}^Lx_n^{s}\right]^{-1}\left[\sum_{l=[L/2]+1}^Lx_l^{1-s}\right]\left[\sum_{n=0}^{[L/2]}x_{n}^s\right]^2,
\label{tausprime}
\ee 
and we obtain the upper bound $\epsilon_1<1/\sqrt{\tau_s'}$. 
We can see that both $\tau_1'$ and $\tau_s'$ are expressed in terms of sums of independent random variables of the form
\be
S_{\sigma}=\sum_{n}x_n^{\sigma},
\ee
with $N=O(L)$ terms and $\sigma$ is either $s$ or $1-s$. 
The large-$N$ behavior of such a sum is well known \cite{gk,bg} to depend on the exponent $\mu$ characterizing the large-$y$ tail of the distribution of $y\equiv x^{\sigma}$, $\rho(y)\sim y^{-1-\mu}$. Using the distribution of couplings in Eq. (\ref{pl}), this exponent is expressed as 
\be
\mu=\frac{1}{2D\sigma}.
\label{mu}
\ee  
For $0<\mu<1$, the expected value $\overline{y}$ is infinite, and the typical value of $S_{\sigma}$, defined as $[S_{\sigma}]_{\rm typ}=\exp\overline{\ln S_{\sigma}}$, is in leading order proportional to $N^{1/\mu}$: $[S_{\sigma}]_{\rm typ}\sim N^{1/\mu}$. For $\mu=1$, $\overline{y}$ is still infinite, and $[S_{\sigma}]_{\rm typ}\sim N\ln N$. For $\mu>1$, $\overline{y}$ is finite, and both the typical and mean values are proportional to $N$: $[S_{\sigma}]_{\rm typ}\sim N$, $\overline{S}_{\sigma}=\overline{y}N$.
We can see from Eq. (\ref{tau1prime}) and Eq. (\ref{tausprime}) that the typical values of $\tau_1'$ and $\tau_s'$ are in leading order proportional to 
a product of typical sums
\be
[\tau_1']_{\rm typ}\sim [\tau_s']_{\rm typ} \sim [S_{s}(L)]_{\rm typ}[S_{1-s}(L)]_{\rm typ}.
\ee
As a consequence, the leading order $L$-dependence of the typical gap must be the same as that of the following combination of typical sums: 
\be 
[\epsilon_1]_{\rm typ}\sim \frac{1}{\sqrt{[S_{s}(L)]_{\rm typ}[S_{1-s}(L)]_{\rm typ}}}.
\ee
Using this relation, Eq. (\ref{mu}), and the known $L$-dependence of typical sums, we find for the $L$-dependence of the typical gap:
\be
[\epsilon_1]_{\rm typ}(L)\sim L^{-z}f_s(L)f_{1-s}(L),
\ee
with the dynamical exponent 
\be
z=\max\{Ds,\frac{1}{2}\} + \max\{D(1-s),\frac{1}{2}\}
\label{z}
\ee
and logarithmic factors for special points:
\be
f_{\sigma}(L) = \begin{cases}
\frac{1}{\sqrt{\ln L}} & {\rm if} \quad 2D\sigma=1 \\
\quad 1 & {\rm otherwise.}
\end{cases}
\ee
The lower bound for the dynamical exponent obtained in Ref. \cite{shirai} from an upper bound on the average gap coincides with $z$ in Eq. (\ref{z}), but that treatment does not account for the logarithmic factors. 
For the asymmetric case $s=0$, there is a special point $D=D_c=1/2$, at which the logarithmic correction appears as $[\epsilon_1]_{\rm typ}\sim\frac{1}{L\sqrt{\ln L}}$ and this explains why the numerically estimated dynamical exponents presented in Ref. \cite{hoyos_epl} deviate from the asymptotic value around this point. In Ref. \cite{shirai}, similar deviations of numerically estimated dynamical exponents appear in the case $D=1$ at the symmetric point $s=\frac{1}{2}$, where according to our results $[\epsilon_1]_{\rm typ}\sim\frac{1}{L\ln L}$.  

\subsection{TFIC with uncorrelated randomness}

Finally, we consider the TFIC with independent, identically distributed random couplings and fields and compare the upper and lower bounds with the gap obtained by the SDRG approximation.  
First, we write $\tau_n$ in terms of the cumulative control parameter defined in Eq. (\ref{u}) as
\be
\tau_n=\sum_{l=n}^L\sum_{m=1}^l\frac{1}{h_l^2}e^{2(u_l-u_m)}.
\label{tauvar}
\ee
For uncorrelated disorder, $u_n$ is a random walk in discrete time $n$ and its mean value behaves as $\overline{u_n}\sim n$ in the ferromagnetic phase, $\overline{u_n}\sim -n$ in the paramagnetic phase and $\overline{u_n}=0$ at the critical point. The fluctuations around the average are $O(\sqrt{n})$. 
The energy gap obtained by the SDRG method \cite{fisher} is 
$\epsilon_{\rm RG}=\min_{1\le m<l<L}\left\{\frac{J_mJ_{m+1}\cdots J_{l-1}}{h_mh_{m+1}\cdots h_l}\right\}=\left[\max_{1\le m<l<L}\{\frac{1}{h_l}e^{u_l-u_m}\}\right]^{-1}$. 
Thus the SDRG gap is essentially determined by the maximal difference in $u_n$, 
$u_{\rm max}=\max_{1\le m<l<L}\{u_l-u_m\}$, as $\epsilon_{\rm RG}\sim e^{-u_{\rm max}}$. 
In the ferromagnetic phase $u_{\rm max}\approx \overline{\ln(J/h)}L$, which yields an exponentially closing gap. At the critical point, the maximal difference scales as $u_{\rm max}\sim\sqrt{L}$, resulting in the stretched exponential scaling  $\epsilon_{\rm RG}\sim e^{-C\sqrt{L}}$, where $C$ is an $O(1)$ random variable \cite{young}. In the paramagnetic Griffiths-McCoy phase $u_{\rm max}\sim \ln L$, leading to an algebraic decay $\epsilon_{\rm RG}\sim L^{-z}$ with a non-universal dynamical exponent \cite{griffiths,mccoy,jli}. Beyond the Griffiths-McCoy phase, in the conventional paramagnetic phase, $u_{\rm max}$ is bounded from above and a finite gap opens.   

We can see in Eq. (\ref{tauvar}), that the dominant term in $\tau_1$ is 
$\max_{1\le m<l<L}\{\frac{1}{h_l^2}e^{2(u_l-u_m)}\}=\epsilon_{\rm RG}^{-2}$. Therefore the lower bound is expected to scale with $L$ in the same way as $\epsilon_{\rm RG}$. Furthermore, we can establish that $1/\sqrt{\tau_1}$ is a lower bound also for the SDRG gap: $1/\sqrt{\tau_1}<\epsilon_{\rm RG}$. In fact, numerical results of Ref. \cite{alcaraz} indicate that $\epsilon_{\rm RG}$ exceeds the exact value of the gap. 

Next, let us consider the upper bound $1/\sqrt{\tau_s}$. 
The stationary distribution in Eq. (\ref{stac}) can be written in terms of $u_n$ as
\be
p_n=\mathcal{N}\exp[2(u_{n_{\rm min}}-u_n)], 
\label{pmin}
\ee
where $n_{\rm min}$ denotes the global minimum of $u_n$, $u_{n_{\rm min}}=\min_n\{u_n\}$. Due to the rapid decrease of $p_n$ with $u_{n_{\rm min}}-u_n$, $p_n$ is typically localized around site $n_{\rm min}$ and, consequently, the normalization is $\mathcal{N}=O(1)$. 
Then $\tau_s$ can be written as 
\be
\tau_s=\mathcal{N}\sum_{l=1}^L\sum_{n=1}^l\sum_{m=1}^l\frac{1}{h_l^2}
\exp[2(u_{n_{\rm min}}+u_l-u_n-u_m)].
\ee
The dominant term of this expression is determined by $\max\{u_l-u_n-u_m\}$ under the conditions $1\le l\le L$ and $1\le n,m\le L$. Obviously, the optimal indices $n$ and $m$ must coincide, therefore we look for the following maximum 
\be 
\max_{1\le n\le l\le L}\{u_l-2u_n\}.
\label{max}
\ee
Denoting the indices which optimize Eq. (\ref{max}) by $l^*$ and $n^*$,
the dominant term of $\tau_s$ is then 
\be
\tau_s\sim \exp[2(u_{n_{\rm min}}+u_{l^*}-2u_{n^*})].
\label{taudom}
\ee
In the ferromagnetic phase, $n_{\rm min}\sim 1$, $l^*\sim L$, and $n^*\sim 1$. The dominant term is thus $\tau_s\sim e^{2u_L}$, which leads to the same exponential decrease of the upper bound $1/\sqrt{\tau_s}\sim e^{-u_L}$ (apart from the prefactor) as that of the lower bound. 
At the critical point, the dominant term in Eq. (\ref{taudom}) is in general different from the term related to the SDRG gap. It coincides with $1/\epsilon_{\rm RG}^2$ only if $n^*=n_{\rm min}$, otherwise they are different. Nevertheless, due to the $O(\sqrt{n})$ fluctuations of $u_n$, the upper bound shows the same type of stretched exponential scaling as the lower bound:  $1/\sqrt{\tau_s}\sim e^{-C'\sqrt{L}}$. 
Finally, in the Griffiths-McCoy phase and in the conventional paramagnetic phase $n_{\rm min}\sim L$, $l^*\sim L$, and $n^*\sim L$. Here, $\tau_s\sim O(1)$, yielding an $L$-independent upper bound, which fails to correctly reproduce the algebraic decrease of the gap in the Griffiths-McCoy phase.

\section{Discussion}
\label{discussion}

Based on an exact relationship with the spectrum of a Markov process, we have formulated lower and upper bounds on the lowest energy gap of open transverse-field Ising chains, which are explicit in the parameters of the model and are valid for arbitrary sets of (non-zero) couplings and fields. 
  
In the ferromagnetic phase and at the critical point, both bounds show the same leading finite-size dependence (with different prefactors). In the homogeneous chain, the upper bound reproduces the correct prefactor in ferromagnetic phase, while at the critical point, the prefactor of the lower bound is slightly closer to the exact one. In the gapped paramagnetic phase, both bounds tend to constants in the limit $L\to\infty$ (the lower bound to zero), and the upper bound becomes more and more accurate farther from the critical point.
In the random TFIC with coupling-field correlations, which is critical and which is a relevant model for adiabatic quantum computing, we showed by the help of the bounds that the finite-size scaling of the gap is related to that of sums of independent random variables. Besides the algebraic closing of the typical (as well as the average) gap we revealed the existence of logarithmic corrections at certain special points. 
The relation of the gap to sums of independent random variables also indicates that, in the anomalous region $z>1$, the gap is essentially determined by the smallest coupling present in the sample, the corresponding term dominating either of or both the sums. 

In the case of uncorrelated disorder, both bounds show the same finite-size scaling in the ferromagnetic phase and at the critical point. 
Furthermore the lower bound accounts for the  algebraically vanishing gap in the Griffiths-McCoy phase, while the upper bound fails to reproduce this (giving a non-vanishing limit). 
We found that the lower bound outperformed the upper bound also at the critical point. 
The former is dominated by the term provided by the SDRG approximation whenever the gap vanishes, including the Griffiths-McCoy phase. 
In the upper bound, which contains the sum $\tau_s\sim\sum_ne^{-2u_n}\tau_n$, the effect of the weighting by $e^{-2u_n}$ is to enhance the term $\tau_{n_{\rm min}}$ at the minimum of the cumulative control parameter. Therefore the dominant term in the upper bound coincides with the SDRG term (related to the maximal increase of $u_n$) only if the starting index of the SDRG term is the same as the global minimum position of $u_n$. Otherwise the upper bound is dominated by some subleading increasing segment of $u_n$, which nonetheless, has the same stretched exponential scaling as the leading one.    

The lower bound used in this paper, although it is less sharp than Laguerre's bound, still shows the same finite-size scaling at the critical point as the exact gap and, due to its simplicity may be more appropriate for analytic treatments, as it was demonstrated for the random TFIC with coupling-field correlations. 

Although we formulated the bounds for the transverse-field Ising chain, they apply also to the closely related XY spin chains and free-fermion hopping models on an open chain. Moreover, the bounds are generally valid for the lowest eigenvalue of tridiagonal matrices of the form $T=BB^T$, where $B$ is a bidiagonal matrix with real and non-zero diagonal and subdiagonal elements.

\begin{acknowledgments}
The author thanks F. Igl\'oi, G. Ro\'osz, and J. A. Hoyos for useful discussions. This work was supported by the National Research, Development and Innovation Office NKFIH under Grant No. K128989, by the Ministry of Innovation and Technology, and the National Research, Development and Innovation Office within the Quantum Information National Laboratory of Hungary.
\end{acknowledgments}

\appendix
\section{Linear term of the characteristic polynomial}
\label{C1}

The relationship in Eq. (\ref{c1tau1}) can be shown by rewriting Eq. (\ref{back}) as $MM^TS\tau=S{\bf 1}$, or as $MM^Tx=\alpha$ with column vectors $x=S\tau$ and $\alpha=(\alpha_1,\alpha_2,\dots,\alpha_L)^T$. Then, according to Cramer's rule 
\be
\tau_1=x_1=\det[(MM^T)_1]/\det(MM^T),
\label{cramer}
\ee
where $(MM^T)_1$ denotes the matrix obtained from $MM^T$ by replacing the first column by $\alpha$. On the other hand, using Jacobi's formula for the derivative of a determinant $\frac{d}{d\lambda}\det A(\lambda)={\rm tr}\left[{\rm adj} A(\lambda)\frac{dA(\lambda)}{d\lambda}\right]$, where ${\rm adj} A$ denotes the adjoint matrix of $A$, we obtain 
$C_1=\frac{d}{d\lambda}\det(MM^T-\lambda\mathbb{I})|_{\lambda=0}=-{\rm tr}[{\rm adj}(MM^T)]$. 
Expanding the determinant in Eq. (\ref{cramer}) by the first column one can see that the terms coincide with the diagonal elements of ${\rm adj}(MM^T)$.  
We note that, for constant transverse fields, the explicit forms of the coefficients of the characteristic polynomial were also given in Ref. \cite{knysh}. 

\section{Comparison with an approximative formula}
\label{app:comparison}

In Ref. \cite{itksz} an approximative formula for the lowest gap $\epsilon_1$, which is the smallest positive eigenvalue of $H$ given in Eq. (\ref{Tmatrix}) was used. Here, we provide a slightly different derivation this formula. 
First, an approximation $v_{\rm app}$ of the eigenvector $v_1$ associated with $\epsilon_1$ is determined. We set $h_L=0$ in $H$ (denoted by $H_L$), which results in $\epsilon_1=0$, and determine the odd components of $v_{\rm app}$ recursively from $H_Lv_{\rm app}=0$. Then we set $h_1=0$ and determine the even components of $v_{\rm app}$ in the same way from $H_1v_{\rm app}=0$. Both odd and even components are normalized to $1/2$.
The approximate gap is then constructed as the expected value $\epsilon_{\rm app}=|v^T_{\rm app}Hv_{\rm app}|$. To compare it with the lower bound, it is expedient to introduce the cumulative control parameter      
\be
u_n = \begin{cases} \sum_{m=1}^{n-1}\ln\frac{J_m}{h_m} \quad {\rm for} \quad  n>1 \\
0 \quad {\rm for} \quad n=1,
\end{cases}
\label{u}
\ee
and recast $\tau_1$ in Eq. (\ref{tau}) as 
\be
\tau_1=\sum_{l=1}^L\sum_{m=1}^l\frac{1}{h_l^2}e^{2(u_l-u_m)}.
\label{tau1var}
\ee
In terms of $u_n$, the approximate gap can written as 
\be
\frac{1}{\epsilon_{\rm app}^2}=\sum_{l=1}^L\sum_{m=1}^L\frac{1}{h_l^2}e^{2(u_l-u_m)}.
\ee
Here, the only difference to Eq. (\ref{tau1var}) is that the upper limit of the second sum extends to $L$. As a consequence, we have
\be 
\epsilon_{\rm app}<\frac{1}{\sqrt{\tau_1}}<\epsilon_1.
\ee

\section{Majorization of the quasistationary distribution}
\label{major}

Let us consider the eigenvalue equations $q(T+\lambda_1\mathbb{I})=0$ and 
$pT'=0$, which determine the quasistationary and the stationary distribution, respectively. Introducing the ratios $R_n=q_{n+1}/q_n$ and $R^0_n=p_{n+1}/p_n$, the above linear equations lead to the following recursions for $1<n<L$:
\beqn
R_n=\frac{h_n^2+J^2_{n-1}-\lambda_1}{J_n^2} - \frac{h_{n-1}^2}{J_n^2}\frac{1}{R_{n-1}}, \nonumber \\
R^0_n=\frac{h_n^2+J^2_{n-1}}{J_n^2} - \frac{h_{n-1}^2}{J_n^2}\frac{1}{R^0_{n-1}},
\label{recursion}
\eeqn
with the initial condition $R_1=\frac{h_1^2-\lambda_1}{J_1^2}$ and $R_1^0=\frac{h_1^2}{J_1^2}$.
We will now show by induction that $R_n<R_n^0$ for $1\le n<L$.
The statement is obviously fulfilled for $n=1$, since $\lambda_1>0$. 
Let us now assume that the statement is valid for $n-1$: $R_{n-1}<R_{n-1}^0$. 
Comparing the terms in the right-hand sides of Eqs. (\ref{recursion}), we see 
that $\frac{h_n^2+J^2_{n-1}-\lambda_1}{J_n^2}<\frac{h_n^2+J^2_{n-1}}{J_n^2}$ and 
$-\frac{h_{n-1}^2}{J_n^2}\frac{1}{R_{n-1}}<- \frac{h_{n-1}^2}{J_n^2}\frac{1}{R^0_{n-1}}$, consequently $R_{n}<R_{n}^0$ holds.

Considering the differences of logarithmic probabilities, we have then
\be
\ln q_{n+1}-\ln q_n<\ln p_{n+1}-\ln p_n
\label{diff}
\ee 
for $1\le n<L$. Since both $q$ and $p$ are normalized, $q_1>p_1$ and $q_L<p_L$ must hold. Furthermore, the inequalities (\ref{diff}) obviously imply that there is only one ``crossing point'' $n^*$ of the two distributions, as anticipated in Eq. (\ref{pq}).     

The inequality (\ref{sqs}) can then be easily proved. We can split the difference $\tau_{qs}-\tau_s=\sum_{n=1}^{L}(q_n-p_n)\tau_n$ into a positive and a negative part:
\be
\tau_{qs}-\tau_s=\sum_{n=1}^{n^*}(q_n-p_n)\tau_n + \sum_{n=n^*+1}^{L}(q_n-p_n)\tau_n 
\ee
Due to the monotonicity of $\tau_n$ [see inequalities (\ref{mon})],
we have the lower bounds for the two parts
\beqn
\sum_{n=1}^{n^*}(q_n-p_n)\tau_n &>& \tau_{n^*}\sum_{n=1}^{n^*}(q_n-p_n) \nonumber \\
\sum_{n=n^*+1}^{L}(q_n-p_n)\tau_n &>& \tau_{n^*}\sum_{n=n^*+1}^{L}(q_n-p_n).
\eeqn  
By adding them, we obtain that $\tau_{qs}-\tau_s>\tau_{n^*}\sum_{n=1}^{L}(q_n-p_n)=0$.


\begin{thebibliography}{99}

\bibitem{pfeuty}
P. Pfeuty, Ann. Phys. (Paris) {\bf 57}, 79 (1970).

\bibitem{lsm}
E. Lieb, T. Schultz, and D. Mattis, Ann. Phys. (N.Y.) {\bf 16}, 407 (1961).

\bibitem{perk}
J. H. H. Perk, H. W. Capel, Physica A {\bf 89}, 265 (1977);
J. H. H. Perk, H. W. Capel, and Th. J. Siskens, Physica A {\bf 89}, 304 (1977).

\bibitem{fisherXX}
D. S. Fisher, Phys. Rev. B {\bf 50}, 3799 (1994).

\bibitem{ijr}
F. Igl\'oi, R. Juh\'asz, and H. Rieger, Phys. Rev. B {\bf 61}, 11552 (2000).

\bibitem{fisher}
D. S. Fisher, Phys. Rev. Lett. {\bf 69}, 534 (1992); 
Phys. Rev. B {\bf 51}, 6411 (1995).

\bibitem{luck}
J. M. Luck, J. Stat. Phys. {\bf 72}, 417 (1993).

\bibitem{ikr}
F. Igl\'oi, D. Karevski, H. Rieger, Eur. Phys. J. B {\bf 5}, 613 (1998).



\bibitem{griffiths}
R. B. Griffiths, Phys. Rev. Lett. {\bf 23}, 17 (1969).

\bibitem{mccoy}
B. M. McCoy, Phys. Rev. Lett. {\bf 23}, 383 (1969).

\bibitem{ir_pre_98}
F. Igl\'oi, H. Rieger, Phys. Rev. E {\bf 58}, 4238 (1998). 

\bibitem{ijl}
F. Igl\'oi, R. Juh\'asz, P. Lajk\'o, Phys. Rev. Lett. {\bf 86}, 1343 (2001).

\bibitem{albash}
T. Albash, D. A. Lidar, Rev. Mod. Phys. {\bf 90}, 015002 (2018). 

\bibitem{hauke}
P. Hauke, H. G. Katzgraber, W. Lechner, H. Nishimori, and W. D. Oliver, 
Rep. Prog. Phys. {\bf 83}, 054401 (2020).

\bibitem{kadowaki}
T. Kadowaki, H. Nishimori, Phys. Rev. E {\bf 58}, 5355 (1998).

\bibitem{dziarmaga}
J. Dziarmaga, Phys. Rev. Lett. {\bf 95}, 245701 (2005);
Phys. Rev. B {\bf 74}, 064416 (2006). 

\bibitem{caneva}
T. Caneva, R. Fazio, and G. E. Santoro, Phys. Rev. B {\bf 76}, 144427 (2007).

\bibitem{dziarmaga_rams}
J. Dziarmaga, M. M. Rams, New J. Phys. {\bf 12}, 055007 (2010).  

\bibitem{lucas}
A. Lucas, Frontiers in Phys. {\bf 2}, 5 (2014).

\bibitem{rams}
M. M. Rams, New. J. Phys. {\bf 18}, 123034 (2016).

\bibitem{delcampo}
F. J. G\'omez-Ruiz, A. del Campo, Phys. Rev. Lett. {\bf 122}, 080604 (2019).



\bibitem{kibble}
T. W. B. Kibble, J. Phys. A: Math. Gen. {\bf 9}, 1387 (1976).

\bibitem{zurek}
W. H. Zurek, Nature {\bf 317}, 505 (1985). 

\bibitem{lukin}
A. Keesling, A. Omran, H. Levine, H. Bernien, H. Pichler, S. Choi, R.
Samajdar, S. Schwartz, P. Silvi, S. Sachdev, P. Zoller, M. Endres, M.
Greiner, V. Vuleti\'c, and M. D. Lukin, Nature {\bf 568}, 207 (2019).

\bibitem{king} 
A. D. King, S. Suzuki, J. Raymond, A. Zucca, T. Lanting, F. Altomare, A. J.
Berkley, S. Ejtemaee, E. Hoskinson, S. Huang, E. Ladizinsky, A. MacDonald,
G. Marsden, T. Oh, G. Poulin-Lamarre, M. Reis, C. Rich, Y. Sato, J. D.
Whittaker, J. Yao, R. Harris, D. A. Lidar, H. Nishimori, and M. H. Amin, arXiv:2202.05847 (2022).

\bibitem{amin}
M. H. S. Amin, Phys. Rev. Lett. {\bf 102}, 220401 (2009).


\bibitem{dr}
B. Damski, M. M. Rams, J. Phys. A: Math. Theor.  {\bf 47}, 025303 (2014).


\bibitem{mdh}
S. K. Ma, C. Dasgupta, and C.-K. Hu, Phys. Rev. Lett. {\bf 43}, 1434 (1979); 
C. Dasgupta, S. K. Ma, Phys. Rev. B {\bf 22}, 1305 (1980).

\bibitem{young}
D. S. Fisher, A. P. Young, Phys. Rev B {\bf 58}, 9131 (1998).

\bibitem{jli}
R. Juh\'asz, Y.-C. Lin, F. Ig\'oi, Phys. Rev. B {\bf 73}, 224206 (2006).

\bibitem{im}
F. Igl\'oi and C. Monthus, {\it Phys. Rep.} {\bf 412}, 277 (2005);
Eur. Phys. J. B  {\bf 91}, 290 (2018).

\bibitem{yr}
A. P. Young, H. Rieger, Phys. Rev. B {\bf 53}, 8486 (1996).

\bibitem{itksz}
F. Igl\'oi, L. Turban, D. Karevski, and F. Szalma, Phys. Rev. B {\bf 57}, 11031 (1997).

\bibitem{ir98}
F. Igl\'oi, H. Rieger, Phys. Rev. B {\bf 57}, 11404 (1998). 

\bibitem{knysh}
S. Knysh, E. Plamadeala, D. Venturelli, Phys. Rev. B {\bf 102}, 220407(R)(2020).

\bibitem{alcaraz}
F. C. Alcaraz, J. A. Hoyos, R. A. Pimenta, Phys. Rev. B {\bf 104}, 174206 (2021).

\bibitem{binosi}
D. Binosi, G. De Chiara, S. Montangero, and A. Recati, Phys. Rev. B {\bf 76}, 140405(R) (2007). 

\bibitem{hoyos_epl}
J. A. Hoyos, N. Laflorencie, A. P. Vieira, and T. Vojta, Europhys. Lett. {\bf 93}, 30004 (2011).

\bibitem{getelina}
J. C. Getelina, F. C. Alcaraz, and J. A. Hoyos, Phys. Rev. B {\bf 93}, 045136 (2016).

\bibitem{gh}
J. C. Getelina, J. A. Hoyos, Eur. Phys. J. B {\bf 93}, 2 (2020).

\bibitem{shirai}
T. Shirai, S. Tanaka, Ann. Phys. {\bf 435}, 168483 (2021).

\bibitem{itr}
F. Igl\'oi, L. Turban, H. Rieger, Phys. Rev. E {\bf 59}, 1465 (1999).

\bibitem{juhasz2022}
R. Juh\'asz, Phys. Rev. B {\bf 105}, 014206 (2022).


\bibitem{redner}
S. Redner, {\it A Guide to First-Passage Processes}, Cambridge University Press (New York, 2007). 

\bibitem{darroch}
J. N. Darroch, E. Seneta, J. Appl. Prob. {\bf 4}, 192 (1967).

\bibitem{doorn}
E. A. van Doorn, P. K. Pollett, Eur. J. Oper. Res. {\bf 230}, 1 (2013).

\bibitem{vankampen}
N. G. van Kampen, {\it Stochastic Processes in Physics and Chemistry}, Elsevier (Amsterdam, 2007).

\bibitem{mg}
C. Monthus, T. Garel, J. Phys. A: Math. Theor. {\bf 43}, 095001 (2010).

\bibitem{gk}
B. V. Gnedenko, A. N. Kolmogorov, {\it Limit Distributions of Sums of Independent Random Variables} (Addison Wesley, Reading, 1954). 

\bibitem{bg}
J.-P. Bouchaud, A. Georges, Phys. Rep. {\bf 195}, 127 (1990).



\end{thebibliography}
\end{document}